**Chaotic uncertainty and statistical inference for natural chaotic systems:**

**Choosing predictors for multiple season ahead prediction of precipitation**

**Extended and Annotated Version 2**

**Michael LuValle**

**Department of Statistics, Rutgers University**

Here we define natural chaotic systems, like the earths weather and climate system, as chaotic systems which are open to the world so have constantly changing boundary conditions, and measurements of their states are subject to errors. In such systems the chaoticity, amplifying error exponentially fast, is so confounded with the boundary condition fluctuations and the measurement error, that it is impossible to consistently estimate the trajectory of the system much less predict it. Although asymptotic theory exists in disconnected pieces for estimating the conditional predictive distributions, it is hard to find where this theory has been applied. Here the theory is reviewed, and applied to identifying useful predictive variables for simultaneous multiseason prediction of precipitation with potentially useful updating possible. This is done at four locations, one midocean three landlocked. The method appears to show promise for fast exploration of variables for multiseason prediction, as well as laying a general foundation for statistical inference for imperfectly understood chaotic systems.

Software instructions for reproducing plot are given in red, first you have to download "Chaotic_uncertainty.pck" to your computer, and use the source() command to load it into vanilla R (not R studio). Then the commands to run are given in red along with some explanation (at first).

The commands first produce a data structure (run time 6 hours to 3 days depending on the problem on a dell laptop running an i9 processor with windows. The algorithm is similar to both bootstrap and random forest, and generates the raw output for subsequent calculations. The second command in the sequence runs an analysis on the data produced in the first part in this case running tests. The same data could be used for prediction intervals as well.

The source code was written in stages, with a lot of commented out code. Some for debugging, others false starts.  For the new data on Davis and Leaburg, download Davis.pck.

**Davissp.nat0full0plus and Leaburgsp.nat0full0plus, correspond to Fsp.nat0full0plus, and Hsp.nat0full0plus in the full download.**

**R packages needed e1071, lars, caret, astsa, parallel**

**Introduction**

The work of Lalley [1] and Judd[2] pointed out that it is not possible to estimate even the path of a chaotic system if measurement error is unbounded (even gaussian), because for any delay map embedding there will be many paths closer to the observations than the actual path. The computational irreducibility theorem, Wolfram[3], and the ability to influence an attractor from a single cell LuValle [4], provide a supporting chorus for this from complex systems theory, particularly for studying computational models of weather and climate and their relationship to actual climate and weather where local phenomena provide disturbances to the attractor. Thus any predictions for real systems need to be multivalued, accounting for the as yet unknown disturbances and mismeasurements. Quite often these distributions are multimodal. With multimodal densities such as these summaries based on the values of the predictive density at the observed value (predictive marginal likelihoods) are more useful measures of prediction than for example Pearson correlation, because they capture the form of the uncertainty in prediction and not just relationships between conditional means and variances. This is important because much of the current literature uses correlation and other mean based tools to choose models, which is analogous to using classical mechanics to analyze the trajectory of electrons in a double slit experiment. Tools for consistent estimation of these predictive marginal likelihoods in chaotic systems are readily available and the paragraph below summarizes an argument for their consistency modulo given regularity conditions. Since my interest is in seasonal prediction, multiple seasons ahead, I will point out that there are some that doubt that seasonal climatic data can be analyzed under the chaoticity hypothesis [5,6] as I do here. So my first example will be an artificial one based on a seaonalized Lorenz attractor with both randomly perturbed tuning parameters and random error in measurement.

The framework that can be derived from Lally's work and Judd's early work is based on an asymptotically consistent estimate of a predictive likelihood for an open chaotic system measured with error using a multiview embedding approach [7,8]. With sufficiently large embedding dimension [9], the set of diffeomorphic embeddings is prevalent [9,10], so local linear regression [11] provides consistent linear prediction for each view within the range of the Lyapunov coefficient for closed chaotic systems with appropriately bounded [1] measurement error. Using Newmann's [12] result combined with Ruelle's [13] on strong mixing of diffeomorphisms of axiom A attractors and an **extension** of the chaotic hypothesis [5,6] to the form of our data (seasonal climate data), the simple kernal and nearest neighbor density and regression estimates [11] of these predictions based on each view (each embedding is a diffeomorphism), will be consistent. Newmann's result remains valid adding in both measurement error (even gaussian) and small random perturbations in boundary conditions that are jointly strongly mixing with the diffeomorphisms (e.g. independent). These predictions are diffeomorphisms implying mixing is exponentially fast[13], so with random sampling of the delay maps across the history we have the capability, for any combination of predictor variables and predicted variable of consistently estimating a predictive distribution. A ***predictive*** likelihood calculated for new observations based on their predictive distribution can be used to compare different sets of predictive variables and on the basis of asymptotic theory to construct tests to compare sets of predictive variables.

In order to capture the time dependence of the predictions, we use the notion of a predictive thread and a predictive tapestry. To create a thread, choose a delay map of sufficient size that it would be diffeormorphic to the attractor [9]. The delay map in the data provides a point in the diffeomorphic prediction space and the j nearest neighbors to that point in the training data are identified. A Y variable (to be predicted) is identified k,k-1, down to 1 season ahead and a regression in the training neighborhood for each delay interval is calculated. The k regressions are applied to the point to be predicted producing a k dimensional vector of predictions from 1 to k seasons ahead. To simulate measurement error for each regression prediction a randomly selected residuals from the training regression is added to it. A single set of predictions plus corresponding random residuals defines a thread. Thousands of such threads are created using randomly chosen delay maps so a tapestry of predictions emerges. At each time point a univariate density is calculated over the threads at that time point. The predictive chaotic log likelihood for that time point is then that density evaluated at the corresponding observed value. The log likelihood for a collection of time points is then calculated as the sum of those predictive log likelihoods . Note the nearest neighbor structure defined by the collection of variables creates a topology which defines the effectiveness and information in this statistical structure.

A statistic corresponding to a paired t statistic can then be calculated by taking the differences between the predictive log likelihood at each point derived from 1 collection of variables minus the predictive log likelihood at each point derived from a different collection of variables (using the same y variable). The variance of the statistic can be calculated via the usual formula for variance of a sum of a time series, and asymptotically the statistic divided by its standard deviation should be distributed as gaussian with mean 0 and variance 1. (here we use a t distribution with simply calculated degrees of freedom to be more conservative) This statistical test of which collection of variables is more informative provides a way to explore the chaotic system for better models especially if supplemented by tying collections of variables back to their mechanistic relations.

In addition the tapestry structure under very particular conditions can afford another, practical advantage in prediction. Under certain conditions a tapestry will learn (only for the extent of time for the tapestry, and note this is not about estimating parameters). As we see new observations along the time period of the thread, we can weigh those threads closer to the observations more than those further away. Under these conditions (not yet theoretically specified) reweighting the density further up in time will improve (increase) the likelihood of the future observed data.

If for example this can work for precipitation, then the tapestry can be the basis for a spreadsheet structure looking forward in time. So by postulating flood in the next time period, we can see how the probability of drought in the following time period changes and evaluate if it is better to release more or less water during a flooding period, or conversly during a particular drought period.

. METHOD OVERVIEW: For each predicted season for each variable we can estimate an initial predictive likelihood, by just estimating the density as the derivative of the weighted empirical distribution function in a small neighborhood of the observed value in the test season. As each season appears, we can reweight the threads for following seasons. Here a gaussian function centered at the new observation is used for reweighting the threads as the data comes in. Inference was done using paired t tests on the log predictive likelihoods in sequence through time, using the auto covariance between the log likelihood differences to account for any cross-year correlation (likelihoods calculated by predicted season), and multiple tapestry creation to account for the variation between random sampling of delay maps.

With this explanation we can now lay out the remainder of this paper. The next section describes the application of the approach to a seasonalized version of the Lorenz attractor. In this section the behavior of the statistic under the null hypothesis is shown to be reasonably well behaved, as is the behavior under a difference between the two collections of predictive variables.. The third shows the application to predicting local seasonal precipitation in two locations, comparing prediction using specific seasonalized space weather vs specific seasonalized earth weather/climate variables, and comparing the latter to a combination of both sets of variables. Although there are many examinations of the influence of solar dynamics on precipitation [eg [14-17]] , this appears to be the first that examines it using statistical methods based on asymptotically consistent methods of predictive likelihood. In addition, it shows that the solar dynamics added to earth based EOF's are particularly more useful in multi season ahead predictions over earth based EOF's, even though the earth based EOF's by themselves dominate in many of the one season ahead predictions.

 The fourth looks at prediction of space weather variables using additional space weather variables vs particular earth climate variables. The last section shows an instance where the learning behavior is working quite strongly, i) for 1 season and one lag in precipitation, and ii) for nearly all seasons and lags for a space weather variable . The results show that the space weather variables may be useful in improving prediction of precipitation and that there is a strong suggestion that earth climate variables may provide a useful probe for helping measure the solar weather cycle.

The software and data used in this is available in the public git-hub repository along with this paper showing how each of the plots in this paper are created.

The algorithm is out lined below:

1) create a "thread" of future histories to incorporate in a tapestry using a randomly chosen delay map of sufficient estimated size so that if the system was closed it would be a diffeomorphism (for practical purposes, 2.5 times the dimension of the manifold estimated by the Levina and Bickel[18] algorithm, as opposed to trying to estimate the box dimension).

2) This delay map in the data to be predicted defines a point in the training data space, and the j nearest neighbors to that point are found.

3) Then a Y variable k seasons ahead, k-1 seasons ahead, … down to 1 season ahead, is chosen for each trajectory in the nearest neighborhood. For each of the k time gaps to the future a regression is calculated in the training data (the j trajectories in the nearest neighborhood) using a Lars[19] algorithm to select a best regression based on Cp [20]. This can be done using training data from climate models [21,22] with further extension of the ideas presented there in or from real data as in this paper.

4) The same regression model is then applied to the test data, and the predicted Y's have a random residual from the training sample added to it to reflect the contribution of measurement errors and continuous varying input conditions.  This creates a series of

predictions+errors for 1 up to k seasons ahead. The predictions are all based on the same nearest neighborhood of trajectories collected from a single fixed delay map so they have a kind of internal consistency with one another for a future prediction.  This defines a "thread." [Note this step requires a little extra if applying regressions created from climate model attractor spinups [21,22] as the residuals need to reflect the error in the true measurements, so there is an intermediate training stage when using climate model data].

5)  Thousands of threads are estimated using a random sampling of possible delay maps of the size defined above. For each of the k time steps in step 4 a predictive density can be calculated, and a predictive likelihood for the kth ahead y variable produced from the test data. These likelihoods for a series of predictions can be calculated across different sets of variables in the simplest approach and pairwise differences can be combined for an asymptotic pairwise test of 0 difference vs one set of variables providing more information than another.

6)  For learning the assumption is then that the internal consistency will **for some of the threads** result in a path consistent with the histories and consistent with the trajectory we are trying to predict. These threads are upweighted and the density recalculated as the local derivative of the empirical distribution function under the reweighting.

7) Seasons are separated as predicted entities, so different models are used for different seasons.

**RESULTS**

**Statistical inference.**

As mentioned above, The inference is based on using paired t (like) statistics calculated by evaluating the predictive likelihood from each predictive point using two different models, or two weightings of the same likelihood (e.g. for the 4 season ahead prediction compare the simple 4 season ahead density at the observation to the 4 season ahead density at the same point using the threads reweighted by the 1 season ahead observation). In both cases there are n years of data, so the n differences are added together and then the variance of that sum of differences is calculated using the usual sum of covariance terms. The ratio of the sum to its standard error is evaluated as a t statistic with n-1 degrees of freedom (more conservative than gaussian) to create a two sided p value. Since there is so much asymptotic theory involved, with significant room for improvement, I am assuming throughout pvalues etc can not be taken at face value but do give a relative indication for exploratory purposes, and all exploratory results need follow up with experimental verification over different time periods. This brings up the difficulty of non-stationarity of the climate system. The results of Albers and Sprott[23] show that for sufficiently high dimension chaotic attractors tend to be continuous. So assuming thedrift in the controlling conditions is slow this provides some support for using the approach despite the non stationarity.

**Plots for displaying results.**

There are two types of inference involved here, one comparing the tapestries created by two different sets of base variables, and the other looking at the learning occurring within a collection of base variables. In all cases in this paper the learning is done using distance of the thread prediction at that time from the true value for the whole collection of variables with all the variables standardized (so each variable has variance one). The weighting for the learning in this paper is using a gaussian with standard error 1 centered at the actual observation.

FDR or False discovery rate [24] plots are created by plotting the ordered pvalues against their ordering (1,2,...,n). The plotted p values can be examined two ways, comparing their distribution to a uniform distribution  (the Null hypothesis distribution, a straight line distribution) and using the false discovery rate algorithm (accounting for between test correlation) to identify discoveries at a .1 false discovery rate level. For each analysis, full tapestries were calculated 5 times so that the between tapestry variation could be assessed. So there will be 5 FDR plots for a given analysis. These will only be shown for the computer experiment.

The other set of plots shows the raw likelihoods in the top (either 2 lines of likelihoods when comparing between two sets of variables or 1 line in the learning case) and the likelihood differences in the bottom plot. In the case of pairwise differences, the bottom plot contains 40 differences, one for 10 calculated predictive likelihoods each season. The 10 are in order, 4 season ahead predictive likelihood, 4 season ahead reweighted by 3 seasons ahead actual value, 4 seasons ahead reweighted by 3 and 2 seasons ahead actual value, and 4 seasons ahead reweighted by 3,2, and 1 season ahead actual value. Then 3 season ahead, with a similar sequence, 2 seasons ahead with a similar sequence and 1 season ahead. In the case of learning there are 6 differences, per season, i.e. unweighted 4 season ahead minus the 4 season ahead with 3 season ahead reweighting etc. The seasons are partitoned using vertical lines. The difference plots also pointwise 95% inference bounds based on the t distributions discussed above. Since there is quite a bit of asymptotically based theory being used here, I follow the practice of providing final conclusions without artificially precise quantification. The four levels of conclusion are. **Not informative** of a difference, **somewhat informative** of a difference (likelihoods showing a consistent direction by season/prediction interval),  **moderately informative** of a difference (showing deviations beyond the .05 bounds), and **highly informative** of a difference (interesting at the false discovery rate level of 0.1 across all 5 replications).

**First experiment, Modified Lorenz model:**

The Lorenz model is calculated using by taking the usual calculation, from time 0 to 50 in .01 time increments, each column is then standardized (subtract mean divide by standard deviation). Then the first and second variables then had independent normal(0,.1) errors added to it while the third  had normal( 0, 1) errors added to it. Then the data was seasonalized by adding the vector 1,2,1,0 repeated 1250 times (with a 1 to cap the end) to each column. Finally every 13th data point was taken until 198 times were included to create a chaotic data set with seasonality some separation and random error, and the same number of data points as our real data set spanning seasons from 1963 through 2012.

The first experiment was done with column 3 being modeled by 3 and 1, vs 3 modeled by 3 and 2 the idea was to test if the results where consistent with there being no real difference in the the models produced in either way. Figure 1a shows the False discovery rate plot for tests the model based on columns 1 and 3, and that based on columns 2 and 3. The pvalue given at the top of the plot is that for a Kolmogorov Smirnov test for departure from a uniform distribution for the p values of the pairwise predictive likelihood tests between the two models calculated for the 10 tests calculated each season.

Figure 1a

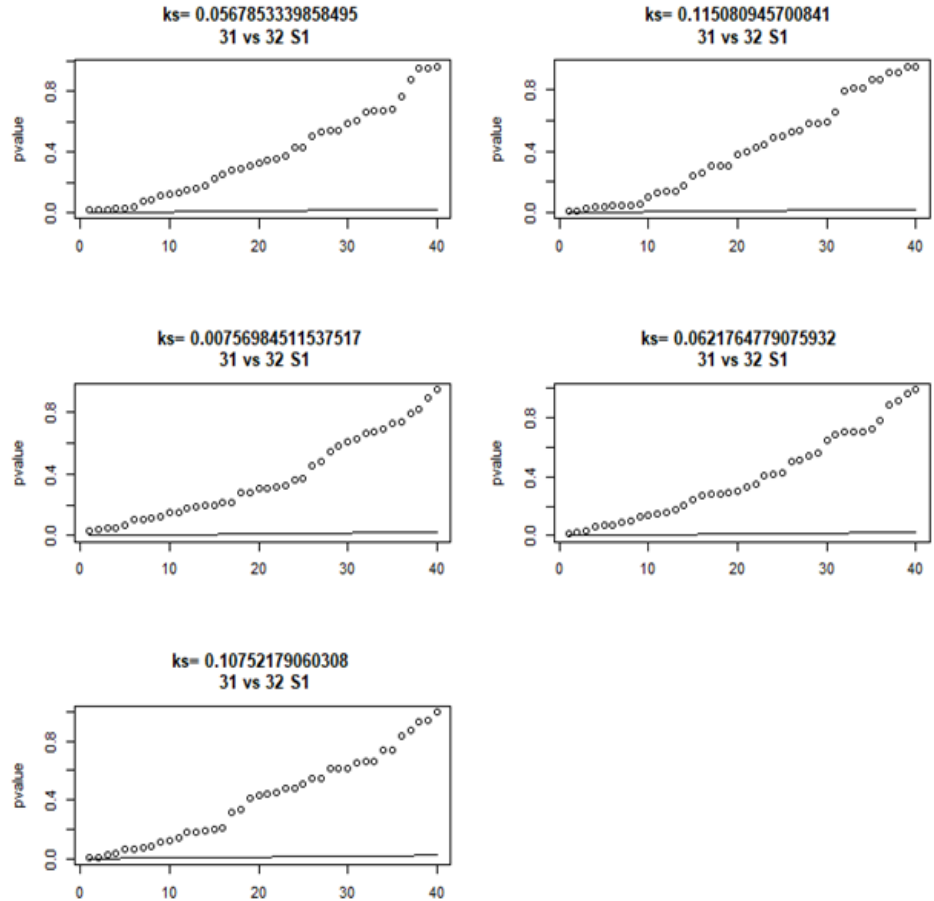

**Legend Figure 1a: Here there are 5 independently created fdr plots (independently sampled from the set of possible delay maps to create the p values). We can see that there is no extreme departure from a uniform distribution (the points run close to a straight line from 0 to 1). The Kolmogorov Smirnov test agrees (although the points have some correlation) so we can not do more than interpret that as a rough similarity measure. None of the points lie below the False discovery rate cut off line in the FDR Plots so none meet a formal false discovery rate criteria of of 0.1.**

Figure 1b

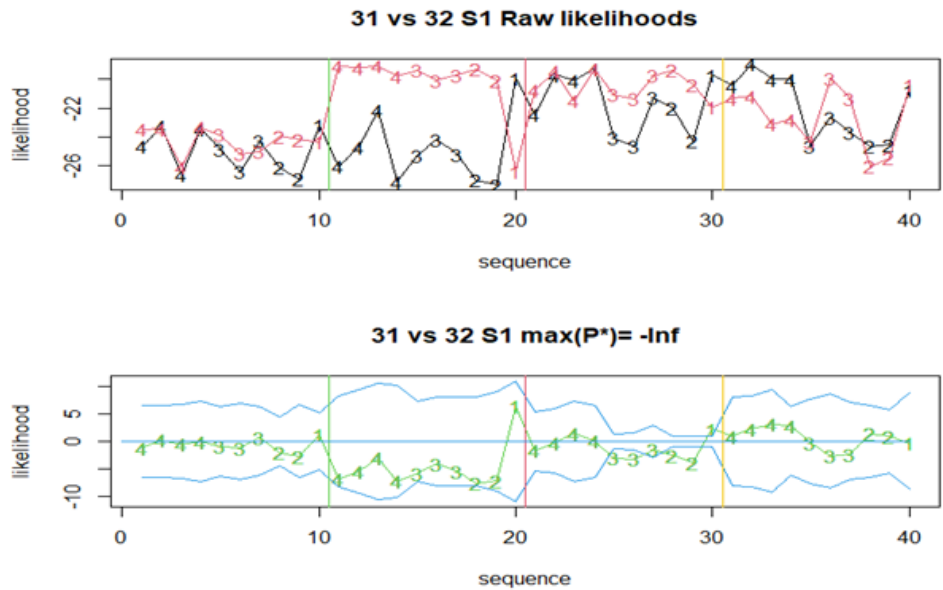

**Legend Figure 1b: The top plot on shows the average "31" likelihoods (black) vs the "32" likelihoods (red) at each point for each artificial season, as generated from all 5 independent delay map resampling. The bottom plot of figure 1 b shows their pairwise differences along with pointwise 95% confidence bounds for the differences.**

<span style="color:red">**Producing the figures**</span>

<span style="color:red">**Data set: LorenzPLUSERRORS1A1**</span>

<span style="color:red">**Command to create likelihoods:**</span>

<span style="color:red">**lorenz.lik31vs32mod3S1Ccopy<-superloop2modL(5,c(3,1),c(3,2),1,1,c(1,2),c(1,2))**</span>

**Function inputs (5=number of random replications of the full random sampling, c(3,1)= variables (columns of data set) in the first model, c(3,2) variables in the second model, 1= variable (from the vector so variable number 3) being modeled in 1st model (1 through 4 seasons ahead), 1=variable being modeled in the 2nd model (likewise), c(1,2)= variables from the selection vector that are included in choosing the neighborhood for learning topology for the 1st model, c(1,2) likewise for the second model. So variable 3 is being modeled in the 1st model, and it is modeled depending on past values of variable 3 and 1 for 1,2, 3 and 4 seasons ahead. The difference in the second model is variable 2 is included as a dependent variable rather than 1.**

<span style="color:red">**Command to create plots:**</span>

<span style="color:red">**FDR: lik.sn.summary2.twomodc("lorenz.lik31vs32mod3S1Ccopy"," 31 vs 32",T)**</span>

<span style="color:red">**Use the data produced above to create FDR plots for likelihood differences at all points using tests based on the described asymptotic theory, FDR plot is set to T.**</span>

<span style="color:red">**Test Plot: lik.sn.summary2.twomodc("lorenz.lik31vs32mod3S1C"," 31 vs 32",F)**</span>

<span style="color:red">**Use the data produced above to create plots for likelihood differences at all points using tests based on the described asymptotic theory, FDR plot is set to F so instead the likelihoods, are plotted in the top plot, the differences with various indications in significance below..**</span>

We see there pvalues are not significantly departing from a uniform distribution and in fact there are no "interesting" tests in any of the replication. Figure 1B shows the actual average log predictive likelihoods overlaid on one another in the top plot with 31 model in black and the 32 model in red. The difference with .05 single point statistical bound is plotted in the bottom plot. In this case the conclusion is that the data is not informative of a departure

Next the modified Lorenz data is modeled predicting column 1 using delay maps from column1 and column 2 vs that predicting column 1 using delay maps from column 1 and column 3. Here the model based on column 1 and column 2 is seen to be consistently better or nearly the same as that based on column 1 and column 3. The use of the model selection accorded by the lars algorithm using minimum Cp model selection may offset some of the poor modeling afforded by column 3. In this case we would say the data is weakly informative of a departure.

Figure 2a

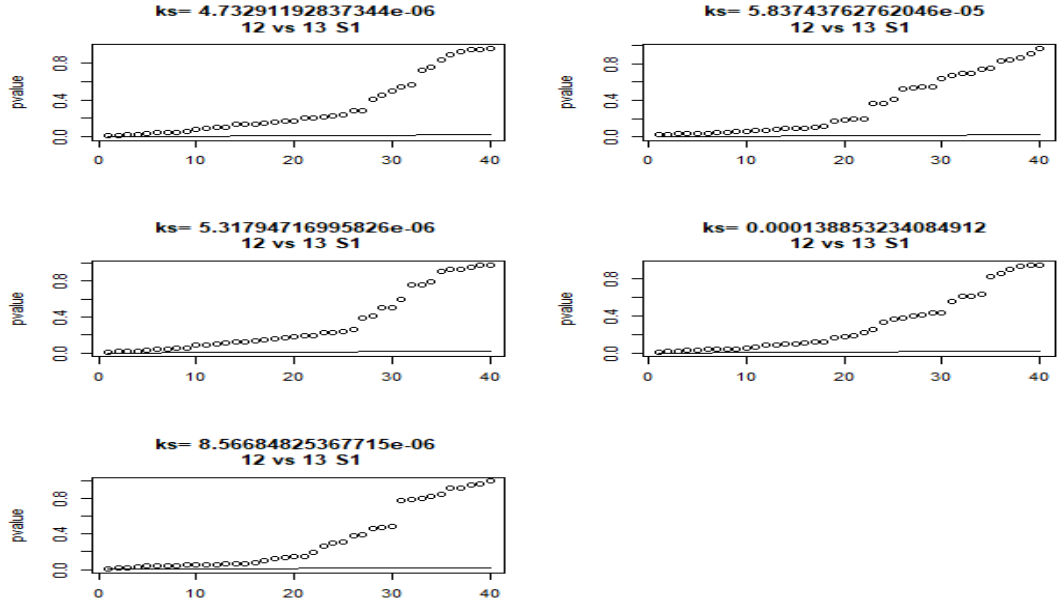

lorenz.lik12vs13mod3S1C

**Legend Figure 2a Figure 2a shows that the p values for tests between the two models depart quite a bit from a uniform distribution**

Figure 2B

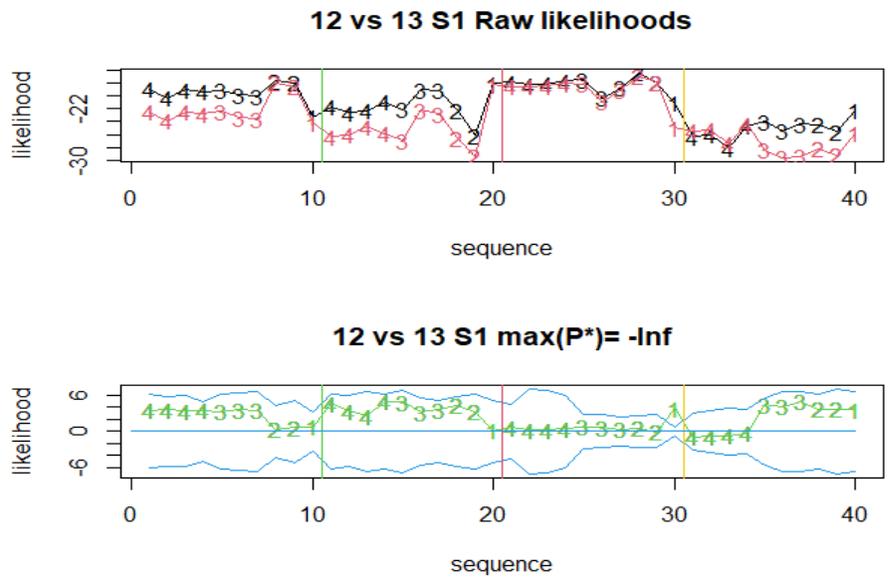

**Legend Figure 2b: Figure 2b shows that the departure is primarily in the direction of the model based on delay maps from column 1 and column 2.**

<span style="color:red">**Command to create likelihoods:**</span>

<span style="color:red">**lorenz.lik12vs13mod3S1Ccopy<-superloop2modL(5,c(1,2),c(1,3),1,1,c(1,2),c(1,2))**</span>

<span style="color:red">**CHECK ABOVE**</span>

<span style="color:red">**Command to create plots:**</span>

<span style="color:red">**FDR: lik.sn.summary2.twomodc("lorenz.lik12vs13mod3S1Ccopy"," 12 vs 13",T)**</span>

<span style="color:red">**Test Plot: lik.sn.summary2.twomodc("lorenz.lik12vs13mod3S1Ccopy"," 12 vs 13",F)**</span>

In Figure 3a below we look at modeling column 1 using delay maps from column 1 and column 2 vs modeling column 1 using delay maps for columns 1, 2, and 3

Figure 3a

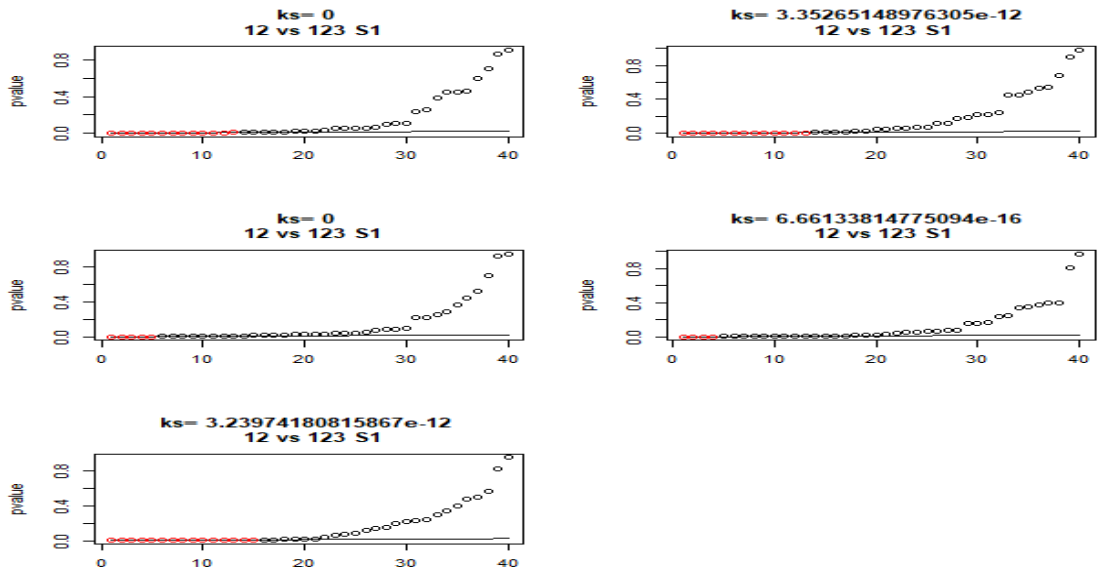

**Legend Figure 3a: Here we see a number of tests (pvalues denoted in red) meet our .1 false discovery rate criterion**

Figure 3b

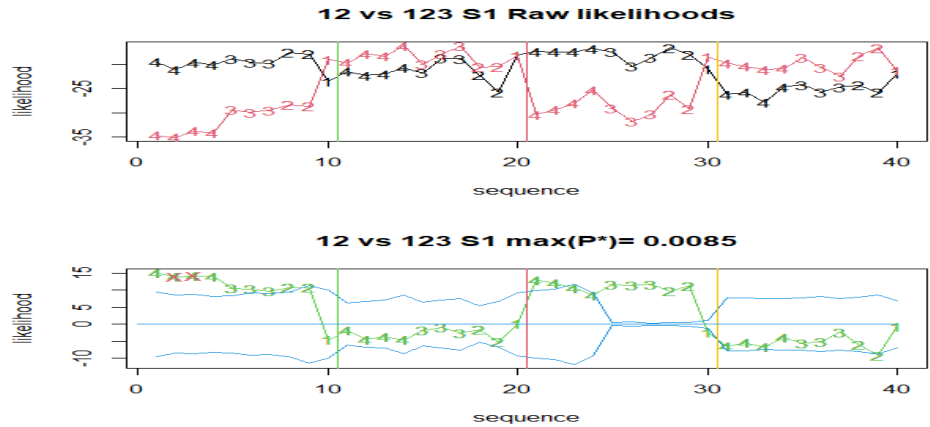

In figure 3b we see a moderate number of points consistently outside the .05 pointwise difference lines while only two are consistently identified by the FDR plots (Red X's). (spring and fall, black is larger than red in the top plot, resulting in a positive difference in the lower plot).

lorenz.lik12vs123mod3S1Ccopy<-superloop2modL(5,c(1,2),c(1,2,3),1,1,c(1,2),c(1:3))

Command to create plots:

FDR: lik.sn.summary2.twomodc("lorenz.lik12vs123mod3S1Ccopy"," 12 vs 123",T)

Test Plot: lik.sn.summary2.twomodc("lorenz.lik12vs123mod3S1Ccopy"," 12 vs 123",F)

Note only 2  of the differences  in spring (both 4 seasons ahead) show red "x"s showing they meet the false discovery rate criterion in all 5 fdr plots. So there is a strongly informative difference for those lags and seasons, but otherwise only a moderate or weakly informative difference.

Finally we see if the framework we've set up promotes learning by testing the points in each season for each lag. A learning sequence would look like the sequence of 4's in the 3$^{rd}$ "season" in section 3, where the predictive likelihoods improve relative to the 1$^{st}$ prediction.

Figure 4a

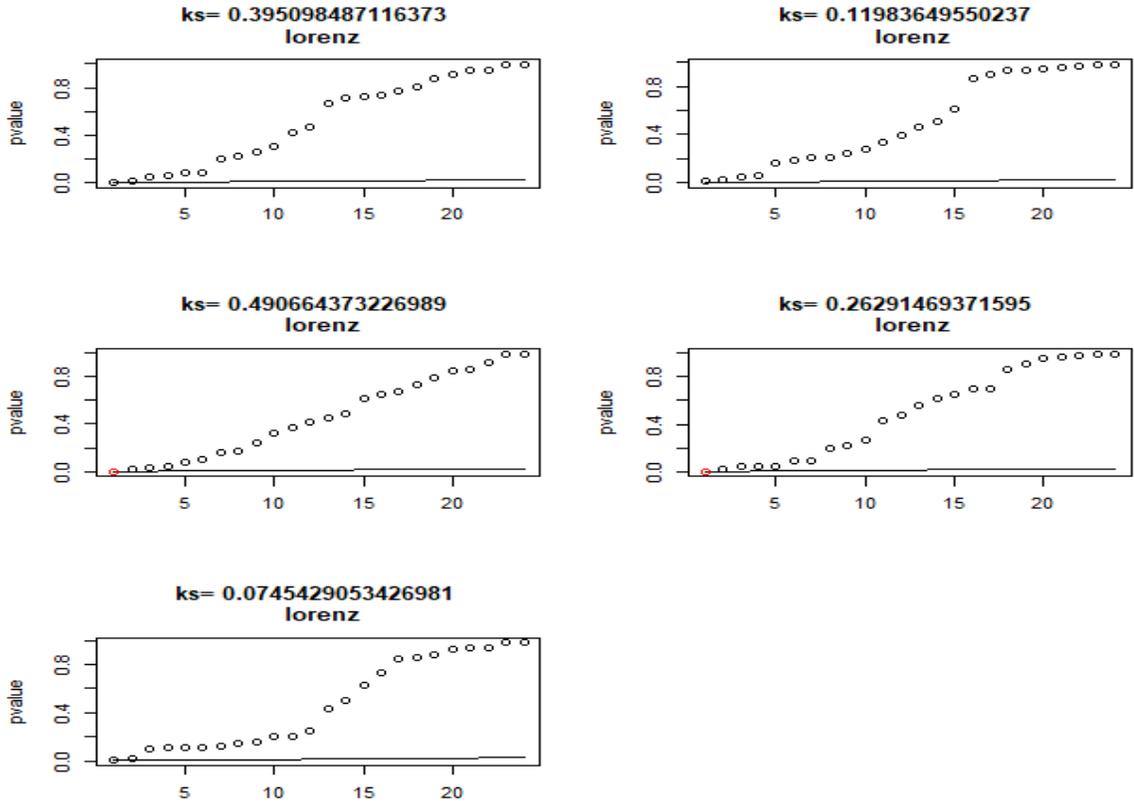

**Legend Figure 4a: Here again we see no real measure of departure from a uniform distribution of p-values consistent with no serious learning taking place.**

Figure 4b

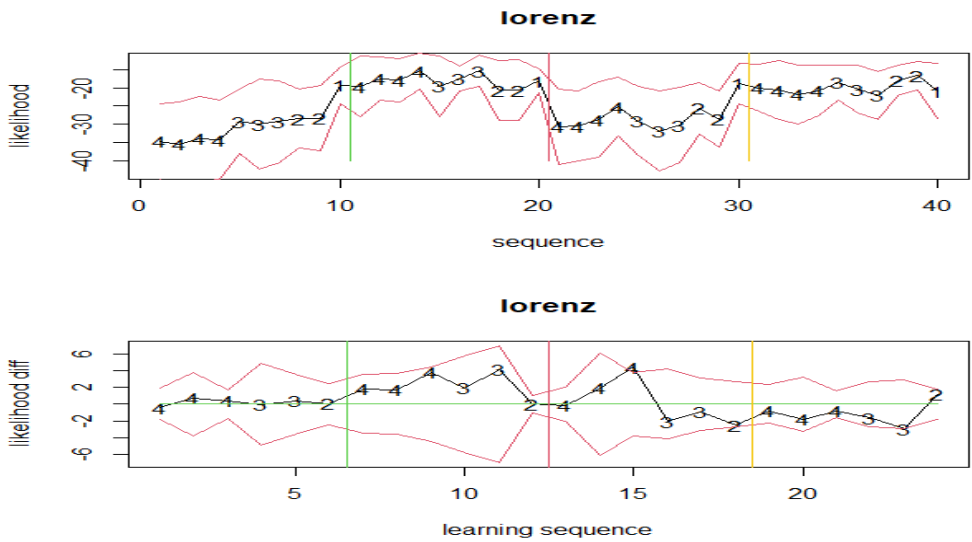

**Legend figure 4b: The upper plot shows the sequence of likelihoods. Examining the summer section, there is the appearance of learning going on among the 4 season ahead and 3 season ahead likelihoods, but the lower plot shows that the learning is not breaking even the .05 pointwise bounds**.

<span style="color:red">liklearnlorenz1<-superloopL(5,1,c(1:3))</span>

<span style="color:red">**In the learning plots, 5 is the number of replications, 1 is the variable being modeled and 1:3 (1 through 3) are being used as the independent variables from the past historyies.**</span>

<span style="color:red">**CHECK ABOVE**</span>

<span style="color:red">**Command to create plots:**</span>

<span style="color:red">**FDR: lik.sn.summary2newL("liklearnlorenz","lorenz",do.plot=T)**</span>

<span style="color:red">**Test Plot: lik.sn.summary2newL("liklearnlorenz","lorenz",do.plot=F)**</span>

Figures 4a and b show that there is no or weak information about learning in this artificial data.

**Rainfall prediction: Comparing models**

In this section we move to real data. The original data included monthly precipitation and temperature from 2 weather stations (one weather station in Fresno California, the other in volcano national park in Hawaii), the Multivariate Enso Index [25], the Pacific Decadal Oscillation index[26], the Arctic Oscillation Index[27], the Indian Ocean Dipole index[28], and three solar activity indexes[29], The Ap index, the sunspot number and the F10.7 index comprising 10.7 cm radiation from the sun reaching earth). For this version data from a weather station in Davis Ca. at the experimental station there, and from Leasburg Oregon have been added and analyzed as well.

Daily and monthly data was converted to seasonal (averaging temperature and the indices, and summing precipitation). It was then standardized to create the data sets in R. Within the analysis program the data in the independent variables is included both with and without seasonal means subtracted. Predictions are made with on data with seasonal means included.

The data matrix has 11 columns. For the Fresno matrix, column 1 is Fresno precipitation, 2 is Fresno Temperature, 3 is the MEI, 4 is the PDO, 5 is the AO, 6 is the IOD, 7 is the Ap index, 8 is the sunspot number, 9 is the F10.7 index, 10 is Hawaii precipitation and 11 is Hawaii Temperature. For the Hawaii matrix, Hawaii precipitation and temperature are columns 1 and 2, and Fresno columns 10 and 11 respectively. The run-to-run variation caused by the random sampling of the delay maps is reflected here in the error calculations for the plots below and can be examined by the reader simply by using the R code provided. In this version another matrix has been created by substituting the Davis precipitation and temperature for Fresno, and Leasburg for Hawaii.

In what follows we will first examine the pointwise difference in likelihood along a tapestry between some different collections of variables that can be used to create predictions. In particular we examine predicting local rainfall measurements in Hawaii and (separately) Fresno, using past local temperature and local rainfall, along with either the three solar variables, or the 4 empirical orthogonal functions (EOF's). In contrast to our Lorenz simulations we see a high signal to noise ratio in the Fresno predictions, and some (moderate) information in the Hawaii predictions. Davis Ca and Leaburg Oregon has been added, chosen because they were: landlocked, off the Pacific (so the same earth climate variables were reasonable) and had fairly good data over the same time frame. Davis was missing a few data points and those were filled in be regressing on Winters Ca data for the same time period.

Figure 5

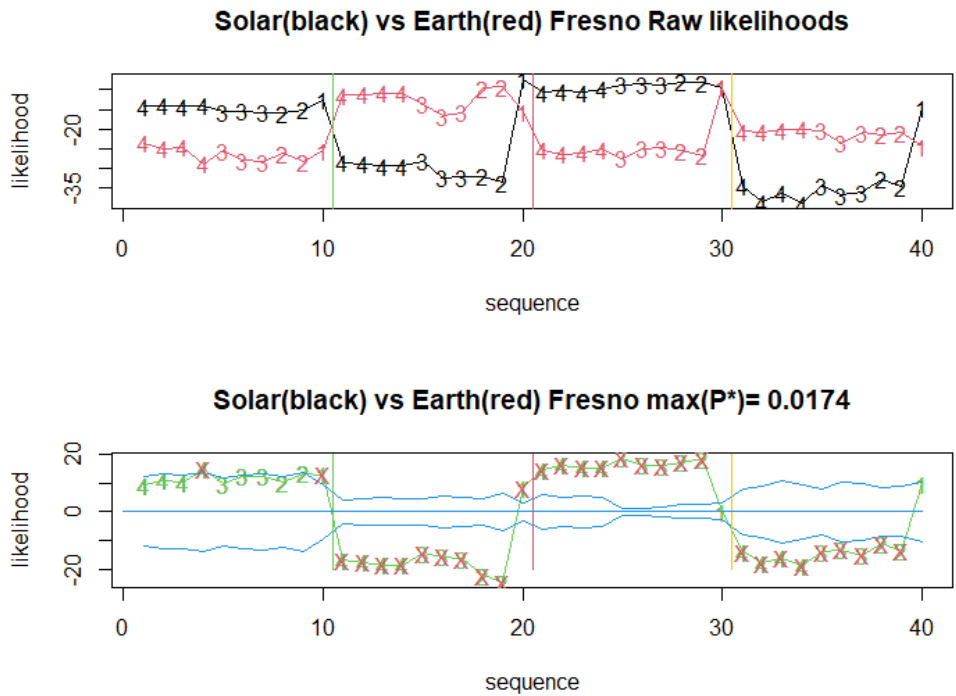

**lik12789vs1to6.copy<-**

**superloop2modSP(5,c(1,2,7:9),c(1:6),1,1,c(1:5),c(1:6),odat1=Fsp.nat0full0plus,odat2=Hsp.nat0full0plus )**

**lik.sn.summary2.twomodcSP("lik12789vs1to6.copy","Solar(black) vs Earth(red)",F,T)**

**Legend Figure 5: Here we see the data is showing a highly informative bias towards the solar based modeling for fall, and toward the earth EOF based modeling for summer and winter (Except perhaps for the 1 season ahead models in a few cases). The spring data is weaker**

Figure 6 Hawaii

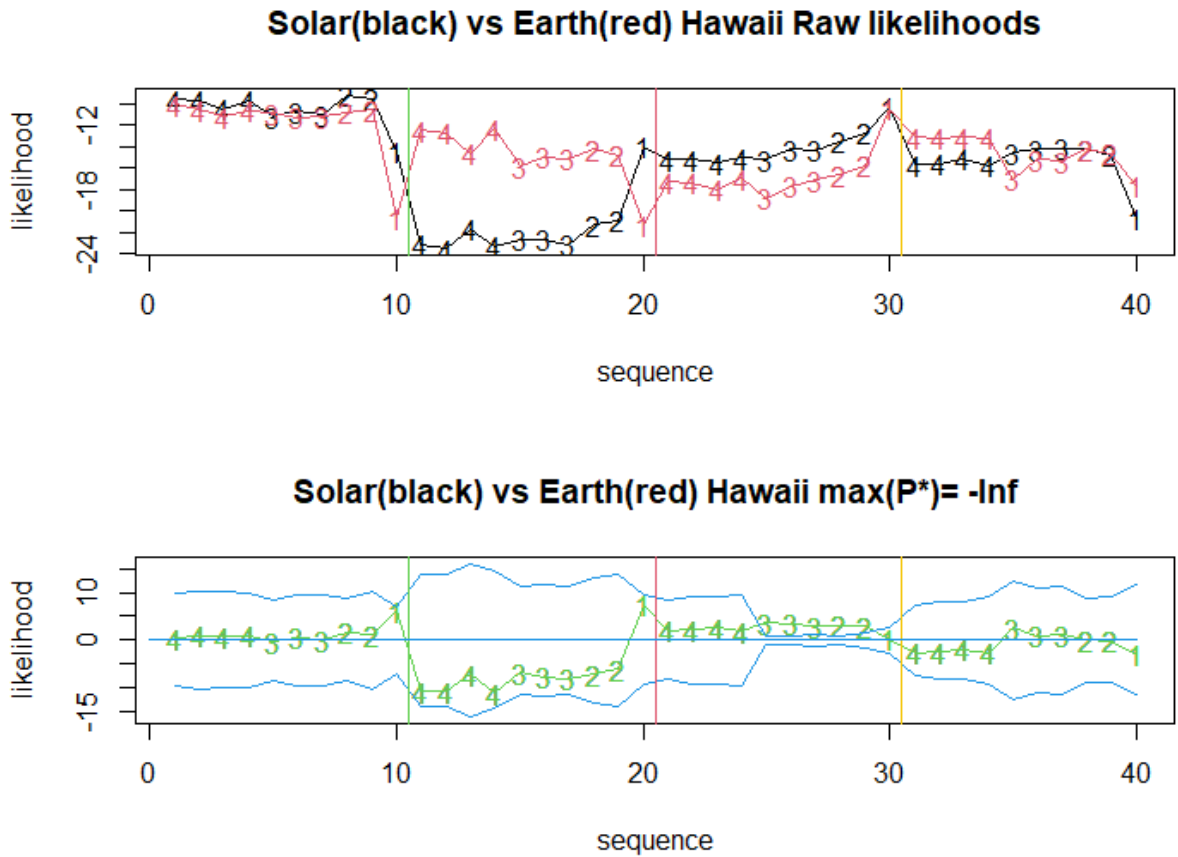

**Legend Figure 6: Here we see a much weaker difference between the two with some dominance of the earth based modeling in summer.**

**Figure 6a  Davis and Leaburg Or**

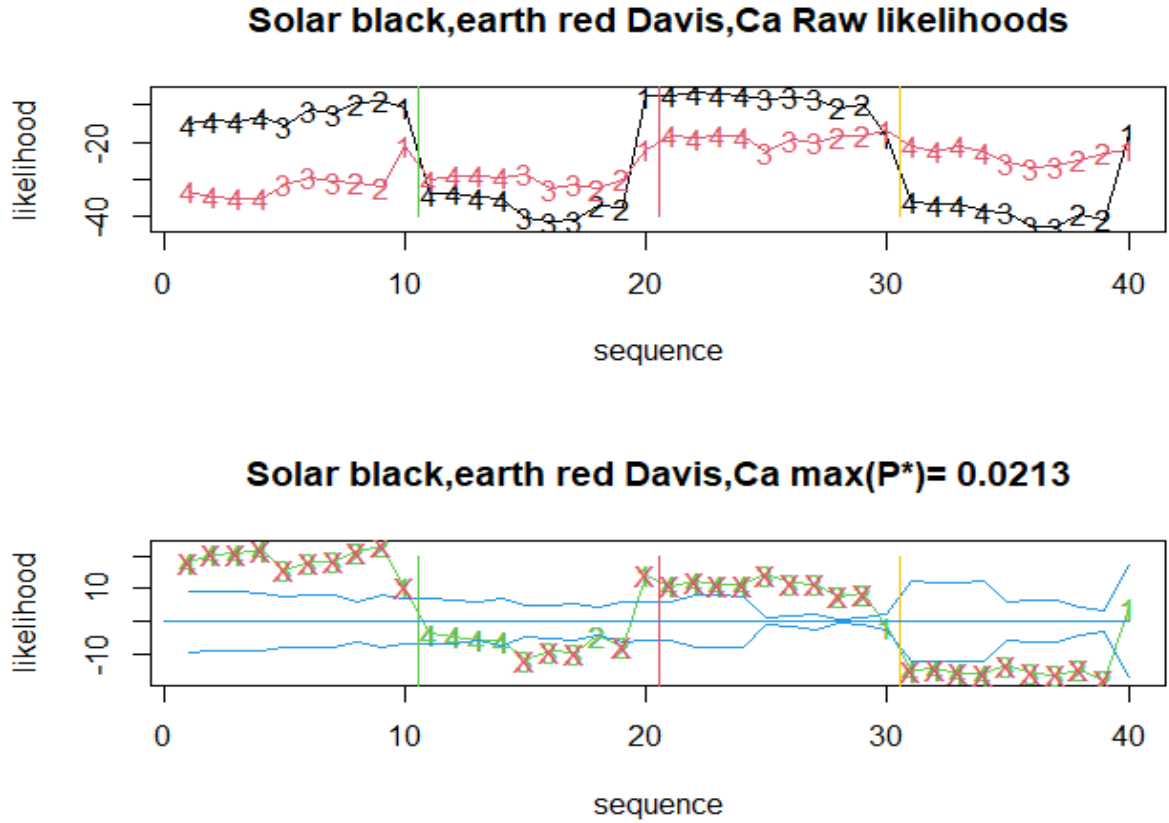

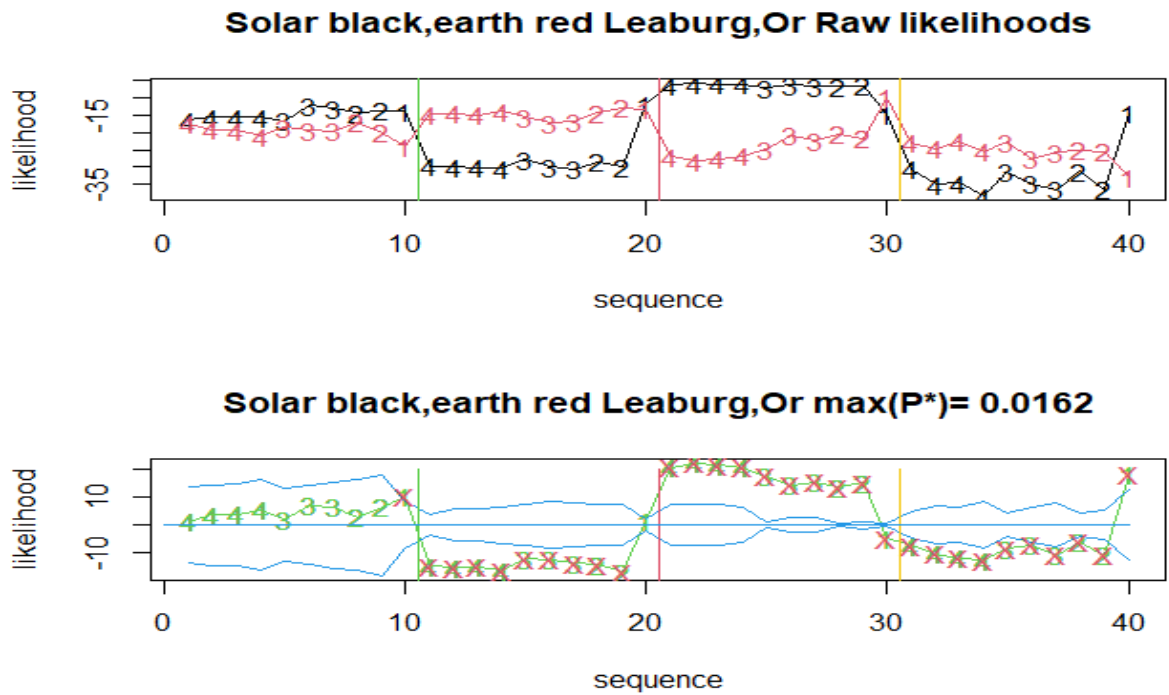

**Legend 6A for these two the pattern is very close to Fresno, the purpose of this was to test the pattern for inland east of the pacific (so the earth models still made sense) and the pattern seems to hold at least for the direct solar measurements vs earthmeasurements as a basis for the information for prediction.**

The results predicted below compare a combination of the solar plus earth variables vs the earth variables along for predicting precipitation in the four locations.

**Figure 7 Davis and Leaburg, Solar + Earth vs Earth**

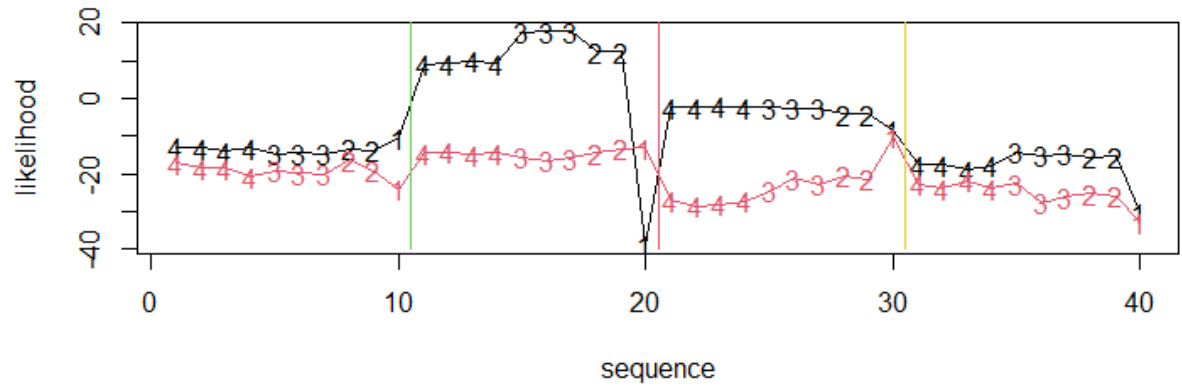

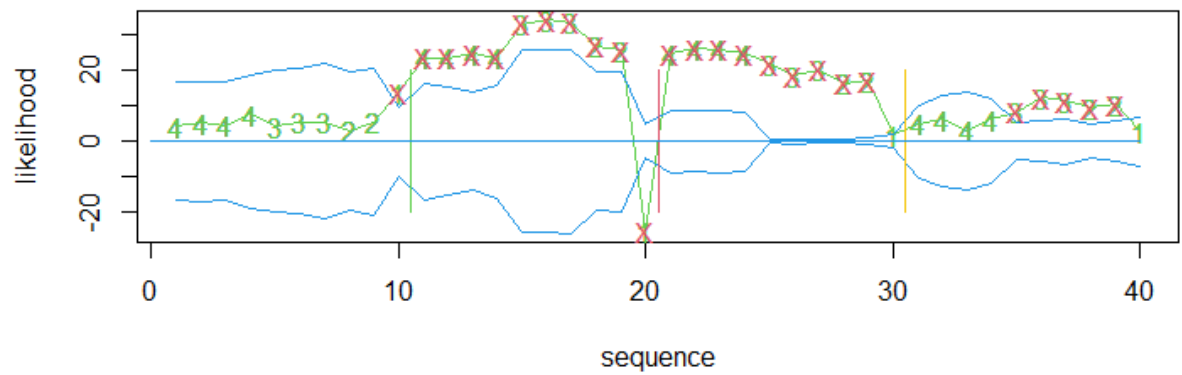

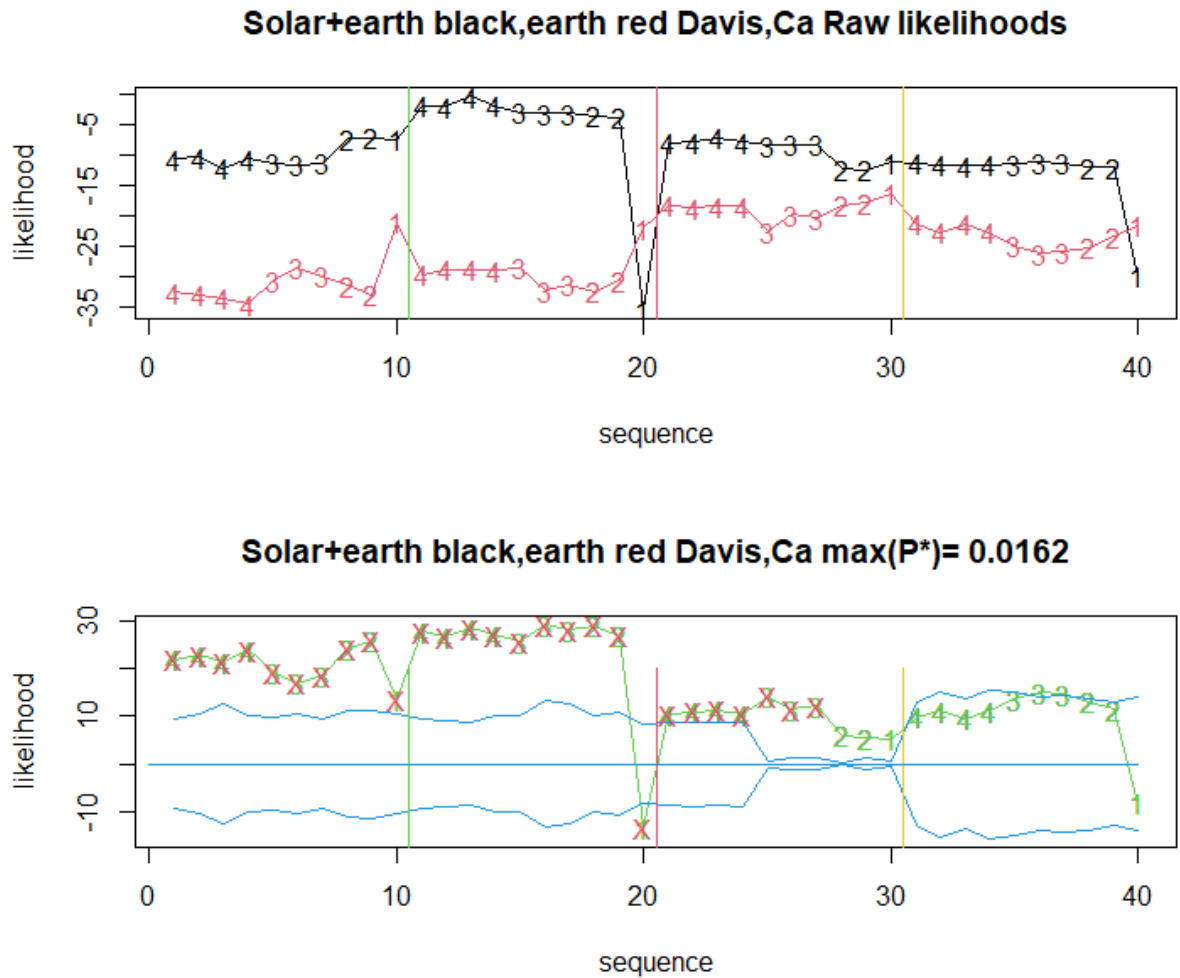

**Legend Figure 7: In both of these again the pattern is similar to that of Fresno, even to earth alone being similar for summer 1 season ahead.**

Figure 8

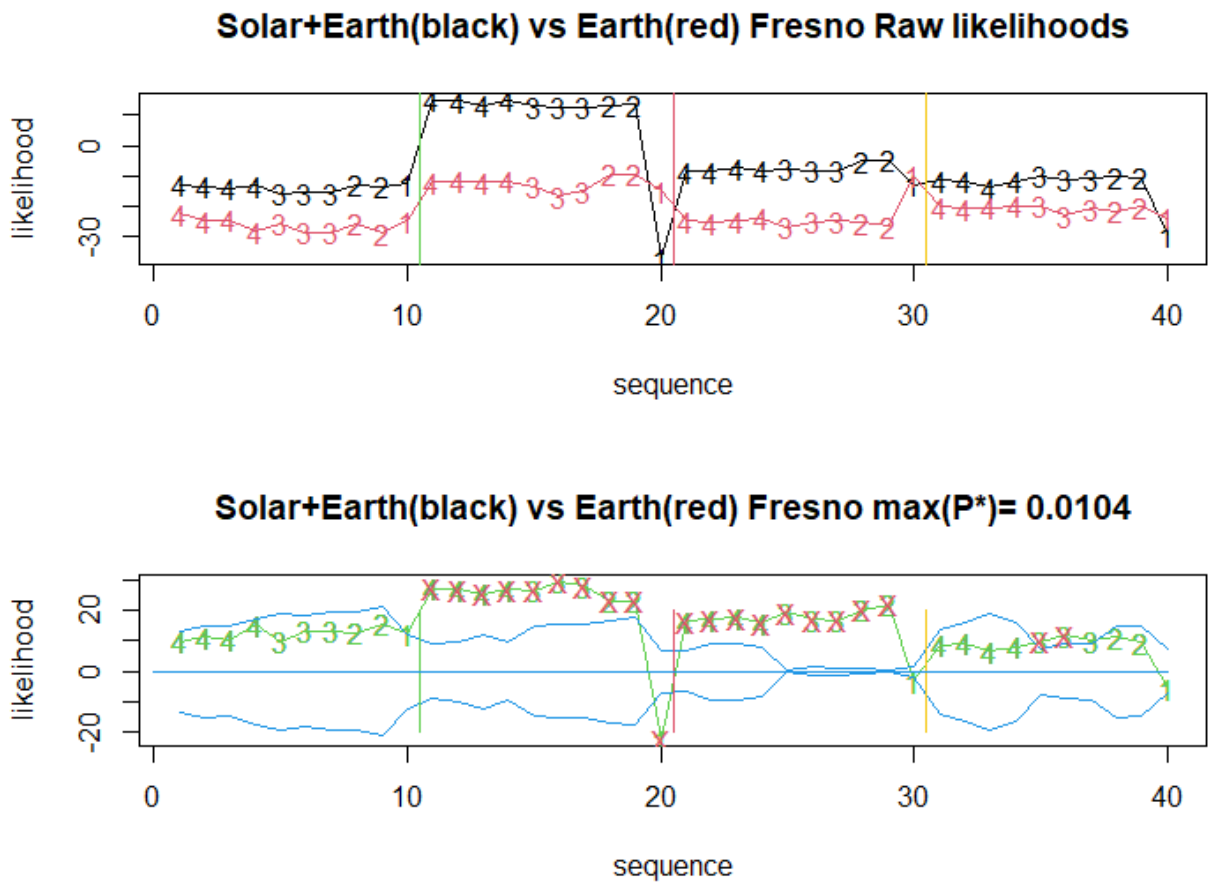

**Legend Figure 8: Here we see that combining solar plus earth seems to be consistently better than purely using earth variables for modeling precipitation in fresno, strongest in summer and fall and quite a bit weaker for winter and spring, and interestingly earth alone seems better for summer fall and winter.**

Figure 9

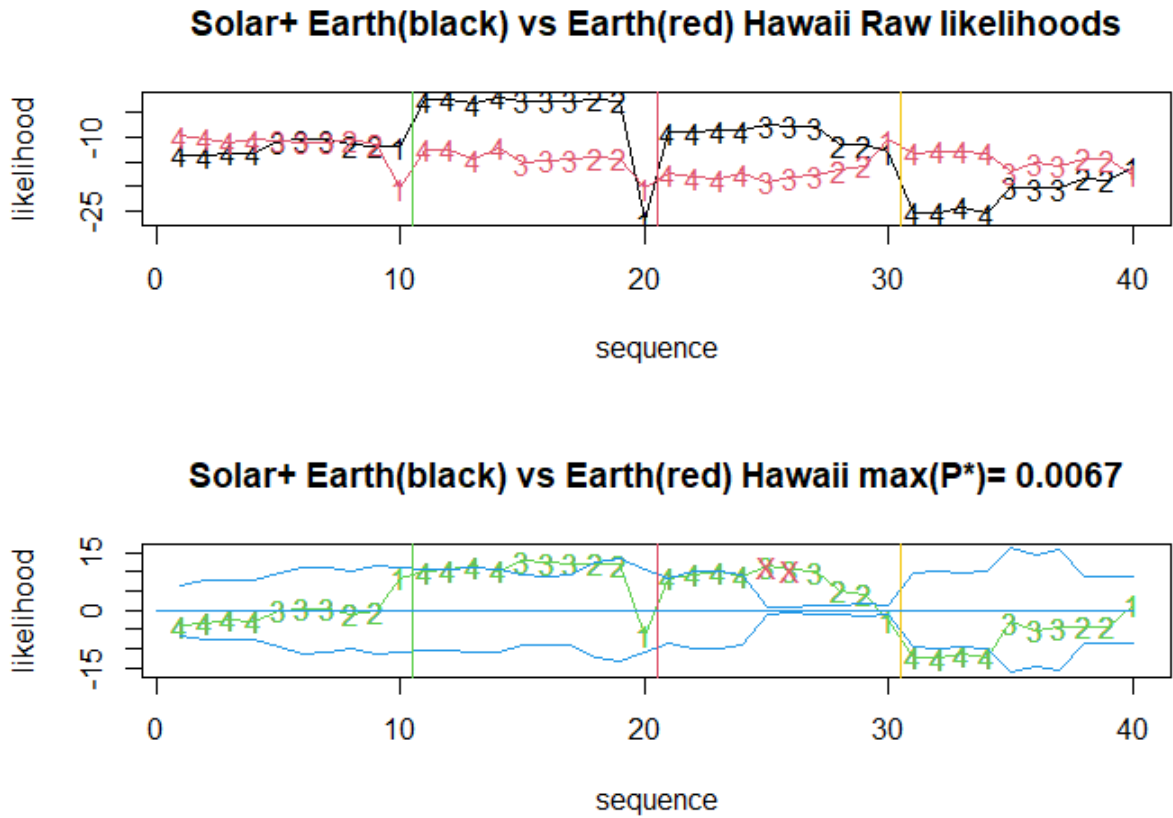

**Legend Figure 9: For precipitation in Hawaii the earth model is weakly dominated by solar plus earth for summer and fall**

lik1to9vs1to6copy<-superloop2modSP(5,c(1:9),c(1:6),1,1,c(1:5),c(1:6), Fsp.nat0full0plus, Hsp.nat0full0plus)

lik.sn.summary2.twomodcSP("lik1to9vs1to6.copy","Solar+ Earth(black) vs Earth(red)",F,F)

[1]

Solar by itself does better for predicting than EOFS by self on spring and fall for Fresno, but worse on summer and winter, For Hawaii, seems to be a wash, either model is reasonable. Solar plus EOF is simply better than EOF except for 1 season ahead in both, though much more significant in Fresno, with only two values in fall passing the FDR criterion for Hawaii.

Conclusion: This suggests strongly including solar dynamics, specifically following geomagnetic disturbances, F10.7 and sunspots, will significantly improve multiseason ahead prediction of precipitation. This can be done using purely empirical models, using a combination of empirical and computational models, or by constructing computational models of Solar dynamics, the solar earth dynamic interface (including biosystems involved) and integrating them into current computational models.

**Predicting the Solar indeces**

In the following 6 plots we compare the use of the three solar indices together in the delay maps for predicting a single solar index, to the use of either fresno or Hawaii temperature and rainfall. The results are consistent with a weak indication of the temperature rainfall information (particularly for Hawaii) containing information that the other solar indeces do not. This shows up mostly in the F10.7 index.

<span style="color:red">Software instructions provided for f10.7, rest are analogous</span>

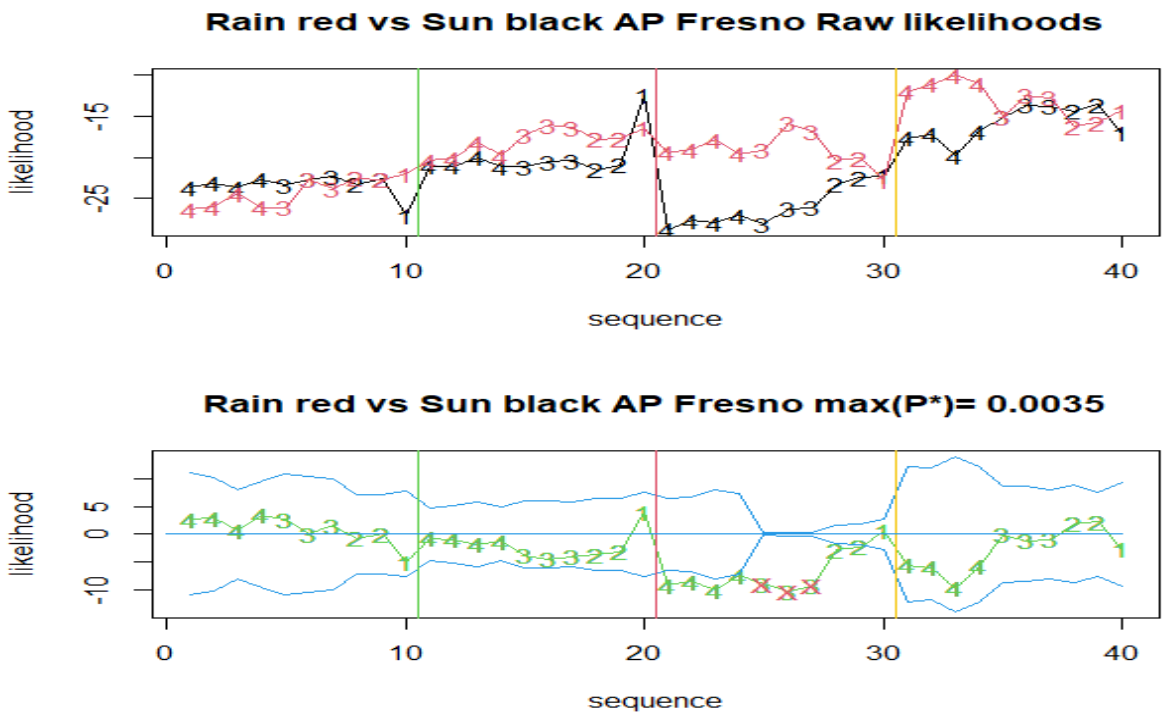

**Rain red vs Sun black AP Hawaii Raw likelihoods**

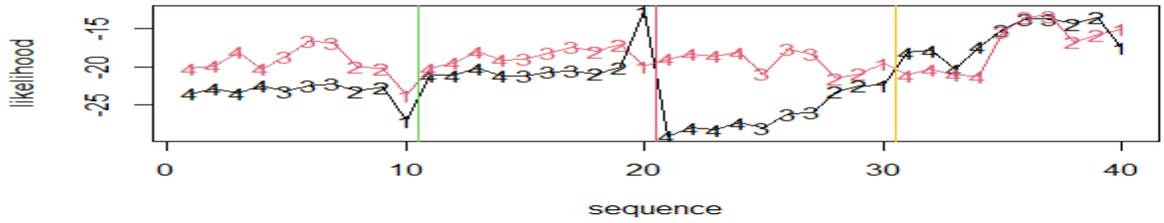

**Rain red vs Sun black AP Hawaii max(P*)= 0.001**

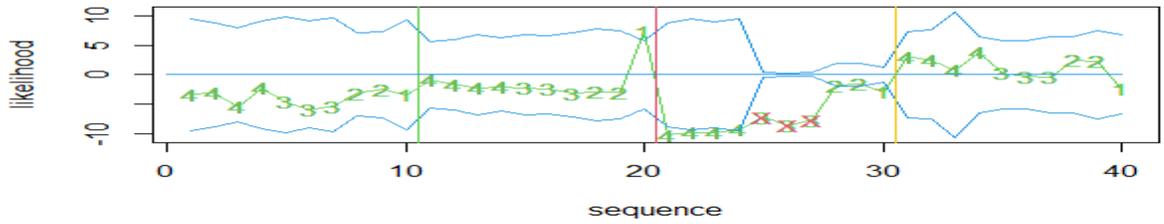

**Rain red vs Sun black SP Fresno Raw likelihoods**

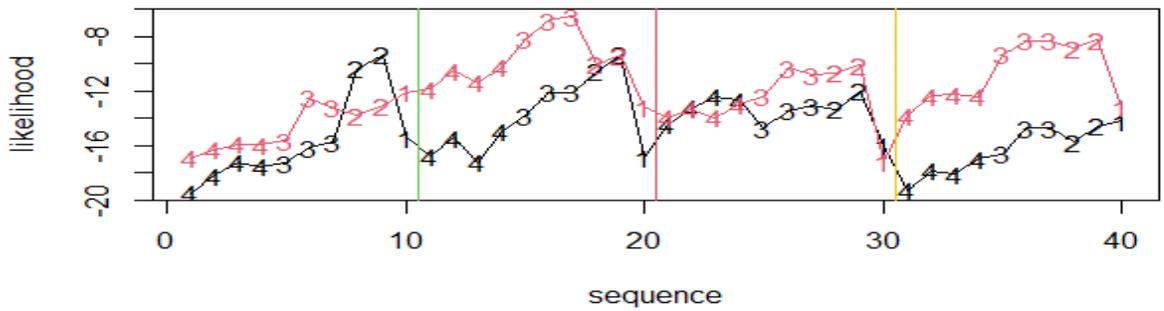

**Rain red vs Sun black SP Fresno max(P*)= -Inf**

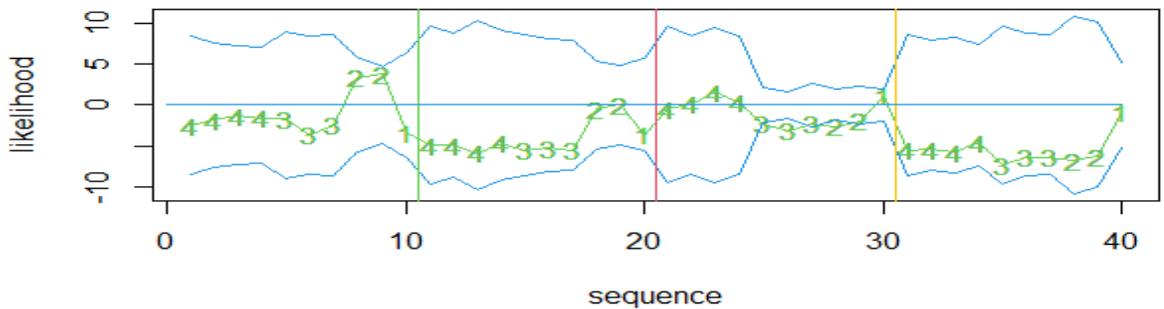

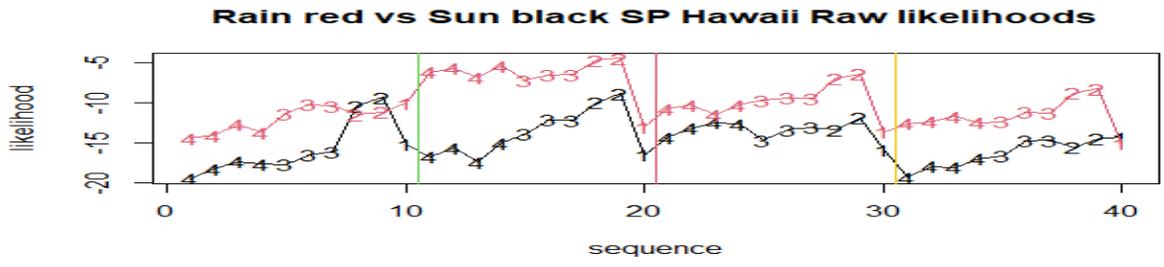

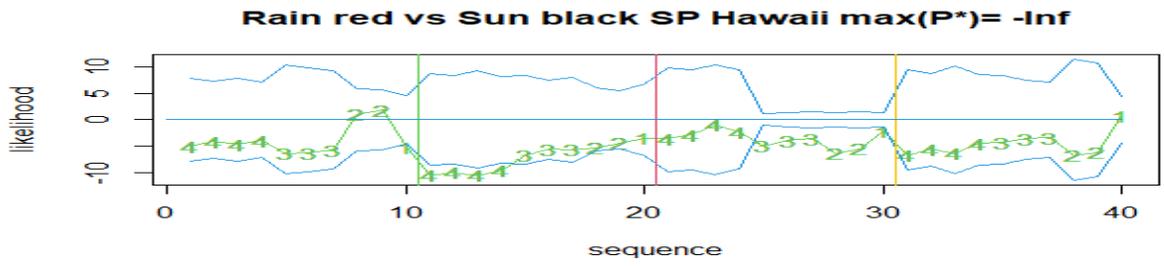

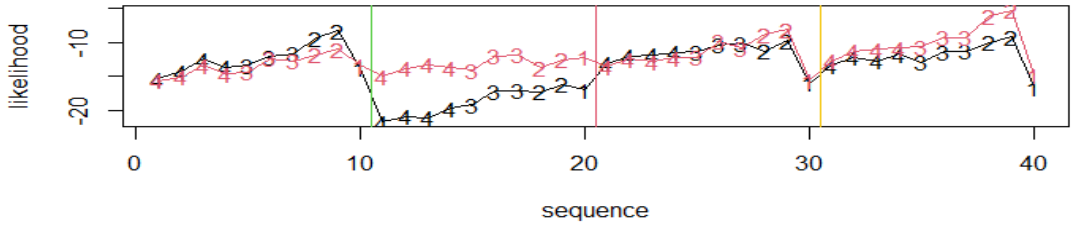

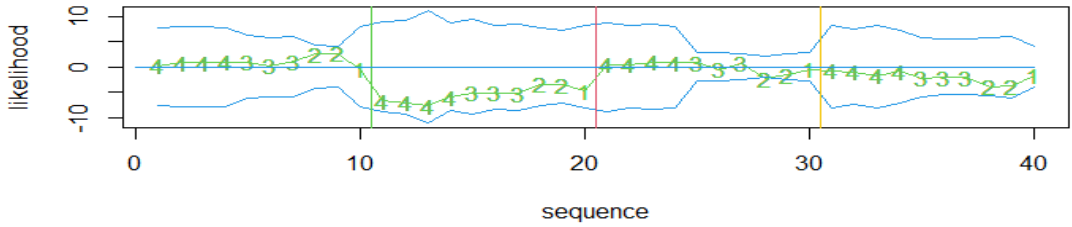

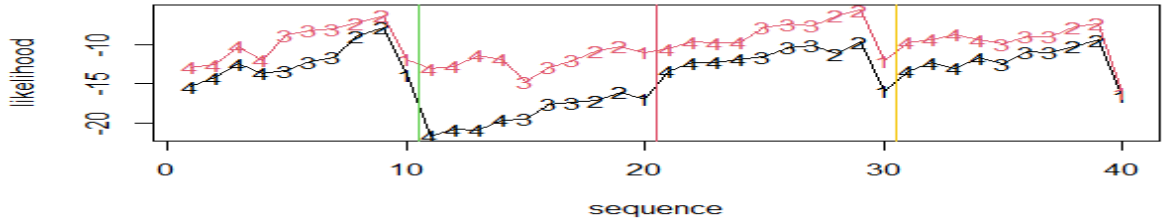

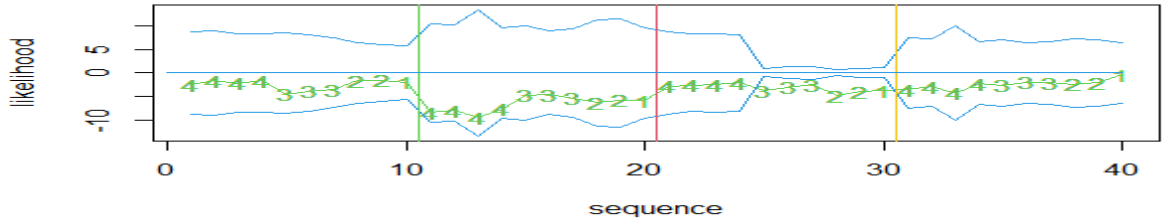

**Lik912vs978,copy<-superloop2modSP(5,c(9,1,2),c(9,7,8),1,1,c(1:3),c(1:3))**

**lik.sn.summary2.twomodcSP("lorenz.lik912vs978mod1S1C","Rain red vs sun black F107",F,T) etc**

There is a general weak indication that earth climate information may provide useful information in terms of modeling solar activity. Possibly because the synchronization of the two systems may be reflecting more information than we are directly measureing from the sun.

.**Learning in the earth sun system**

Two data sets are examined for learning, first prediction of Hawaiian precipitation from itself and Fresno temperature and precipitation, shown in figure 11 and prediction of F10.7 from the F10.7 data, and Fresno temperature and precipitation, shown in figure 12. Learning is assessed by an increase in likelihood after reweighting, vs the unweighted likelihood for each lag for prediction.

For learning in the Hawaiian Precipitation (Figure 11) , there are only 2 points clearly outside the global 95% difference lines in the second plot down. In all four seasons the $1^{st}$ 3 season ahead prediction shows some learning, In summer both 3 season ahead predictions show a high level of learning, passing both FDR tests. In Fall the $1^{st}$ 3 season ahead prediction and the 2 season ahead prediction passes the global FDR criterion. In winter both 3 season ahead predictions are beyond the .05 line.

For the sun cycle prediction of F10.7 (figure 12), nearly every place learning can occur is being picked as positive by all three criteria, the exception being 3 of the 4 season ahead predictions, although 1 of those is outside the .05 pointwise lines.  It is interesting that except for spring, the 1 season ahead prediction based purely on embedding has lower likelihood than any of the other predictions.

**Figure 11: Learning in prediction of Hawaiian Rainfall**

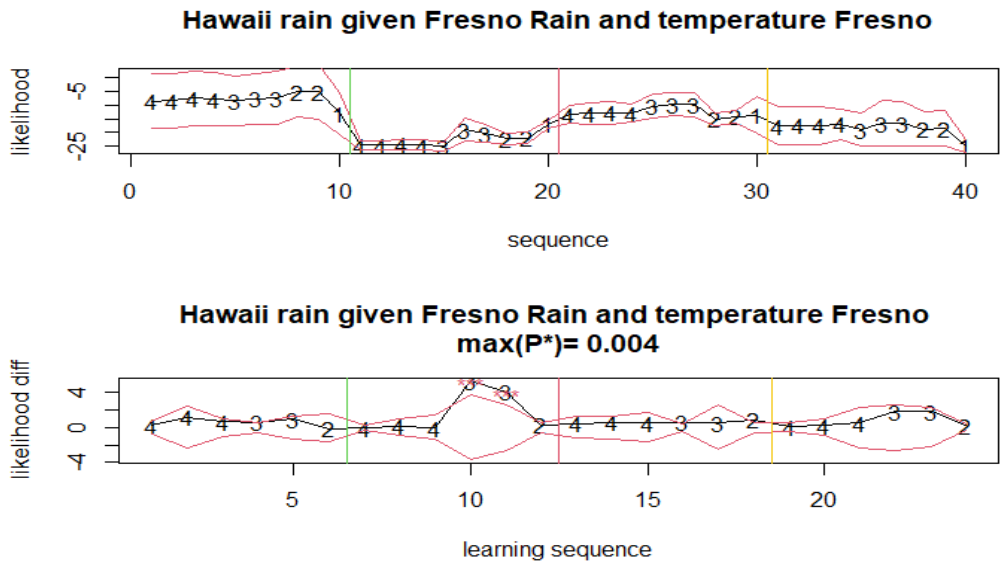

liklearn.312<-superloopSP(5,c(1,2,10),y=3,cvar=c(1:3), ), Fsp.nat0full0plus, Hsp.nat0full0plus)

lik.sn.summary2new("liklearn.312 ","Hawaii rain given Fresno Rain and temperature",do.plot=F,Fresno=T)

Legend Figure 11: Here since the likelihood differences are between the first prediction for a given lag and the subsequent predictions for that lag. Otherwise the notation is similar, and the actual likelihood is shown with only 1 line and a simple set of 95% confidence bounds.

Figure 12: Learning for F10.7 prediction

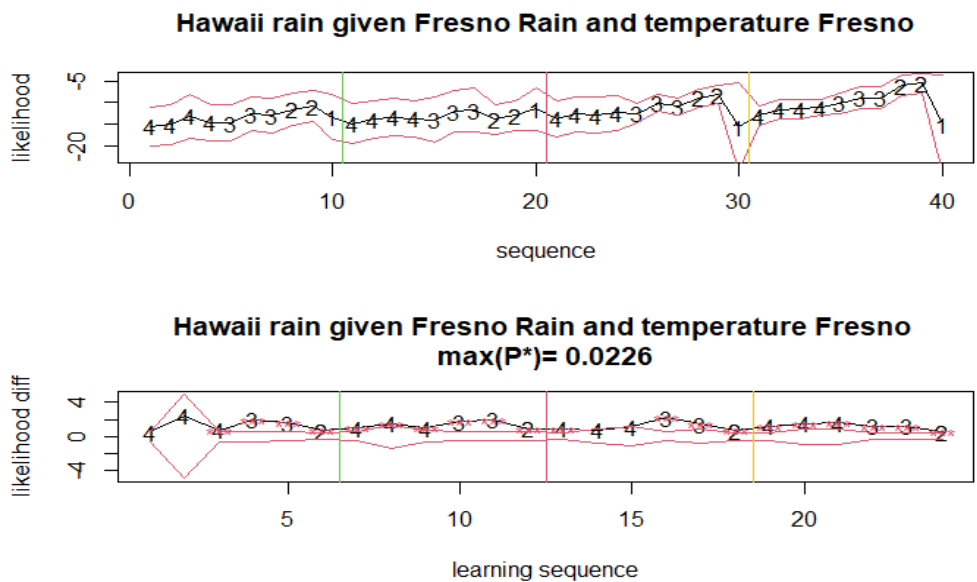

liklearnF107.trial912<-superloopSP(5,c(9,1,2),y=3,cvar=c(1:3), ), Fsp.nat0full0plus, Hsp.nat0full0plus)

lik.sn.summary2new("liklearnF107.trial912","F10.7",do.plot=F,Fresno=T)WORKING ON CORRECTION

Legend Figure 12: Nearly every point is showing some learning.

Discussion and conclusion

An (asymptotically) statistically valid method avoiding the assumption that a conditional predictive distribution given the past has a simple parametric form has been presented, empirically validated against an artificial example, and applied to the problem of identifying better models for predicting precipitation multiple seasons in the future. There is strong evidence that prediction more than one season ahead is dependent on tracking precisely the state and previous states of the solar cycle and the suns various outputs. This is especially true for the landlocked examples though there is some evidence for it in the mid ocean example. Adding the Davis and Leaburg stations strengthens the evidence, but much more remains to be done. One immediate point is that this statistical result points to the earth system having long term memory of the state of the solar cycle. If for example, vegetation is the source of this long term memory, perhaps through the combined effects of multiple solar outputs (e.g. light frequencies ratios through time) it makes the case for even more caution in playing with geo-engineering methods that do not take such into account.

For efforts to improve seasonal forecasts multiple seasons ahead, the result suggests that either a very capable prediction model for variations in the solar cycle is necessary, along with concommitment improvement of understanding the mechanisms (e.g. plant-based mechanisms) by which precipitation is influenced by the sun over long lead times without effecting the usual EOF's, or a far better approach to hybrid computational statistical models than that offered by the two references [25,26] needs to be developed. The statistical approach developed here can be extended to test which kinds of such models would be best at any given time.

The learning result for precipitations seems strong for only one season and one lag, though there are several further avenues to explore with the learning concept for statistical tapestries in natural chaotic systems, the most obvious being the weighting function for checking learning. The sporadic performance for only a few lags could be due either to the interaction of randomness and chaoticity or be analogous to effects in purely chaotic systems as observed in the physical literature [30] for embedding models. Learning for the solar cycle seems quite strong perhaps because there is significantly less noise in the measurement of the target variable?

The test data here is only the last 9 years of the data sets, for two locations. Now updated to four. Clearly more validation is necessary for this to be useful. The programs and data used here are available on Git hub at  (https://github.com/2718LuValle/Chaotic-Uncertainty ). Someone familiar with R should be able to extend this easily to other data sets and time intervals

The method does not depend on the application being climate. Any application for which a reasonable generalization of the chaoticity hypothesis holds, and in which the systems has either random boundary conditions and or is measured with error should be amenable to this approach. Examples include confined plasma flow in reaching toward fusion power, or more speculatively in flows of social information and activity using different approaches to social media and news communication.

Any commercial application of this approach involving the use of global climate models should be done after reviewing the patents [21,22] whose current owner is OFS® Fitel. The author would receive no compensation for such use.

## References and Notes

1) Lalley, 1991, "Beneath the noise, chaos" Annals of Statistics 1999, V27, #2, 461-479

2) Judd, 2007, "Failure of maximum likelihood methods for chaotic dynamic systems", Phys. Rev. E, 75 039210

3) Wolfram, S. **A New Kind of Science**, 2002 Wolfram media, ISBN: 1-57955-008-8

4) LuValle, B. "Chaotic Manipulation of Weather Systems using Hybrid Cellular Automata Techniques", https://symposium.foragerone.com/meeting-of-the-minds - 2024/presentations/65319

5) Bonetto, F., Galavotti, G., Giuliani, A. and Zamponi, F. (2006),"Chaotic Hypothesis,Fluctuation Theorem and Singularities", Journal of Statistical Physics.V123, no 1. DOI: 10.1007/s10955-006-9047-5

6) Galavotti, G. (1996), "Chaotic Hypothesis: Onsager Reciprocity and Fluctuation-Dissipation Theorem", Journal of Statistical Physics, V84, No5/6, 899-925

7) Ye, H., and Sugihara G.,(2016), "Information leverage in interconnected ecosystems, overcoming the curse of dimensionality", Science, Vol 353, issue 6502, 922-925

8) Deyle E.R. Sugihara, G (2011), Generalized Theorems for Nonlinear State Space Reconstruction, PLoS ONE 6(3), e18295, doi:10.1371/journal.poine.0018295

9) Sauer, T., Yoreck, J. and Casdagli, M., 1991 "Embedology", Journal of Statistical Physics, 65, 579-616

10) Hunt, Sauer, York, 1992, "Prevalence, a translation invariant "almost every" on infinite dimensional spaces", Bulletin of the AMS, 1992, v27, n2, October 1992, 217-238.

11) Charles J. Stone, *The Annals of Statistics*, Vol. 5, No. 4 (Jul., 1977), pp. 595-620
https://www.jstor.org/stable/2958783

12) Neumann, M.H., (1998),"Strong approximation of density estimators for weakly dependent observation by density estimators from independent observations", Annals of Statistics, V26, no 5, 2014-2048

13) Ruelle, D, (1976), "A measure associated with axiom A attractors", American Journal of Mathematics, V98, no 3, 619-654

14) Thomas, Vinceth, and Abraham, (2024), "Solar activity and extreme rainfall over Kerala india", arXiv:2047 18262v1

15) Zhang, L., Liu, Y., Zhan, H., Jin, M., and Liang,X., 2021, "Influence of solar activity and El Nino-Southern Oscillation on precipitation extremes, streamflow variability and flooding events in an arid-semiarid region of China, Journal of Hydrology 601, 126630

16) Nytka, W. and Burnecki, K.,(2019), "Impact of solar activity on precipitation in the United States", Physica A, 121317

17) Nesme-Ribes,E., Fiere, E.N., Sadourny, R., Le Treut, H., and Li, Z., X., (1993), " Solar Dynamics, and its impact on total solar irradiance and the Terrestrial Climate", Journal of Geophysical Research, V98, no A11, pp 18923-18935, 1993

18) Levina, E. and Bickel, P. J. (2005). "Maximum likelihood estimation of intrinsic dimension". Advances in NIPS 17. MIT Press

19) Trevor Hastie and Brad Efron (2013). Lars: Least Angle Regression, Lasso and Forward Stagewise. R package version 1.2. https://CRAN.R-project.org/package=lars

20) Mallows, C. L. (1973). "Some Comments on $C_P$". *Technometrics*. **15** (4): 661–675. doi:10.2307/1267380. JSTOR 1267380

21) LuValle, M., 2016, "Predicting climate data using a climate attractors derived from a global climate model" US Patent 9262723, Assignee OFS Fitel

22) LuValle, M., 2019, "Statistical Prediction functions for natural chaotic systems and computer models thereof", Patent 10234595, Assignee OFS Fitel

23) Albers, and Sprott, 2006, "Structural Stability and Hyperbolicity violation in High Dimensional Dynamic Systems", Nonlinearity 19, 1801-1847

24) Benjamini and Y. Hochberg, "Controlling the false discovery rate: A practical and powerful approach to multiple testing," J. Roy. Statist. Soc. Ser. B, no. 57, pp. 289–300, 1995.

25) Wolter, K., and M. S. Timlin, 2011: "El Niño/Southern Oscillation behaviour since 1871 as diagnosed in an extended multivariate ENSO index" (MEI.ext). *Intl. J. Climatology*, **31**, 14pp., 1074-1087. DOI: 10.1002/joc.2336.  (DATA https://psl.noaa.gov/enso/mei.ext )

26) Mantua, N.J. and Hare, SR (2002), "The Pacific Decadal Oscillation", Journal of Oceanography, Vol. 58, pp. 35 to 44, 2002 (DATA https://www.ncei.noaa.gov/pub/data/cmb/ersst/v5/index/ersst.v5.pdo.dat )**NEED tO CHECK**

27) Higgins, R. W., A. Leetmaa, Y. Xue, and A. Barnston, 2000: Dominant factors influencing the seasonal predictability of U.S. precipitation and surface air temperature. *J. Climate*, **13**, 3994-4017. (DATA https://www.cpc.ncep.noaa.gov/products/precip/CWlink/daily_ao_index/monthly.ao.index.b50.current.ascii.table )

28) Behera, Swadhin K.; Yamagata, Toshio (2003). "Influence of the Indian Ocean Dipole on the Southern Oscillation". Journal of the Meteorological Society of Japan. 81 (1): 169–177. (Data: **This is new, my data was originally from an Australian site need to check** https://psl.noaa.gov/gcos_wgsp/Timeseries/DMI/ )

29) Matzka, J., Stolle, C., Yamazaki, Y., Bronkalla, O. and Morschhauser, A., 2021. The geomagnetic Kp index and derived indices of geomagnetic activity. Space Weather, *https://doi.org/10.1029/2020SW002641*   ( DATA https://kp.gfz-potsdam.de/en/data )

30) Garland J. and Bradley E.,  "Prediction in Projection", Chaos **25**, 123108 (2015); doi: 10.1063/1.4936242

31) TEMPERATURE AND PRECIPITATION DATA FROM: Climate Data Online ( https://www.ncei.noaa.gov/cdo-web/  mapping tool  Hawaii station USC00511303, Fresno station USW00093193, Davis station USC00042294, Leaburg station USC00354811)

**Acknowledgments:** I would like to thank Professors Han Xiao, Rong Chen, and John Kolassa for discussion and references to Neumann's article. And I would like to thank Doug

Nychka whose comment on an earlier version helped me find a software error that resulted in false regularity in the data. In addition I would like to thank Luke Beebe, Kavi Chakkapi, Yvyn Vyas, Prisha Bhamre, and Anastasya Chuchkova, whose discussion helped clarify many issues.